\documentstyle[aps,amssymb,12pt]{revtex}

\begin{document}
\author{Charles Picciotto}
\address{Department of Physics and Astronomy, University of Victoria, Victoria,\\
British Columbia, Canada V8W 3P6}
\title{$K^{\pm }\rightarrow \pi ^{\mp }\mu ^{\pm }\mu ^{\pm }$ and doubly-charged
Higgs}
\date{February 1997}
\maketitle

\begin{abstract}
The rate for the lepton-number-violating decay $K^{\pm }\rightarrow \pi
^{\mp }\mu ^{\pm }\mu ^{\pm }$ is calculated in a model which incorporates
doubly-charged Higgs bosons. \ We find that for reasonable values of the
parameters the decay branching ratio may be as large as $2\times 10^{-16}.$
\ Although this is a discouragingly small number, it is of the same order of
magnitude as the rate mediated by massive Majorana neutrinos.
\end{abstract}


\section{INTRODUCTION}

In the past couple of decades a large amount of effort has been dedicated to
the study of double beta decay in nuclei. \ The motivation has been
primarily two-fold: \ The experimental observation and analysis of the
two-neutrino process $(A,Z)\rightarrow (A,Z+2)+2e^{-}+2\bar{\nu}$ has shed
light on the structure of the nuclear matrix elements involved, and the
search for the neutrinoless process $(A,Z)\rightarrow (A,Z+2)+2e^{-}$ has
provided a powerful tool for the study of the Majorana properties of
neutrinos, with considerable speculation about what might be inferred from
the actual observation of this lepton-violating decay.

The studies of neutrinoless double beta decay have mostly relied on the
assumption that the process would be mediated by massive Majorana neutrinos,
with the modern picture being described by leptons which appear in left- and
right-handed doublets, the large mass of the right-handed neutrino being
responsible for the small left-handed neutrino mass via the seesaw
mechanism. \ In this scheme neutrino double beta decay would proceed as a
second-order process, described by the same Hamiltonian that is responsible
for the standard weak processes.

However, the possibility that double-beta decay might be mediated by a
different lepton-violating interaction was also considered early on\cite
{Smith}, with the suggestion that a $\triangle _{3/2,3/2}$ resonance inside
the nucleus might induce a lepton-violating process. \ Later, a mechanism
for $|\Delta L|=2$ was proposed where the electrons might couple to quarks
via a doubly-charged Higgs\cite{Vergados}, but subsequent arguments showed
that in this case the effective coupling is negligibly small\cite
{Wolfenstein}. \ In fact, the conclusion for nuclear double beta decay was
that there are no important $|\Delta L=2|$ contributions not proportional to 
$m_{\nu }$ in a $SU(2)\times U(1)$ model with a Higgs triplet.

The Majorana nature of neutrinos can in principle also be tested in the
double beta decay of strange mesons, although one would expect very small
branching ratios. \ Experimental searches for nuclear double beta decay put
stringent limits on $K^{\pm }\rightarrow \pi ^{\mp }e^{\pm }e^{\pm }$. \
Calculations for $K^{\pm }\rightarrow \pi ^{\mp }\mu ^{\pm }\mu ^{\pm }$
have been performed in a model where lepton-number violation is transmitted
through the propagation of a heavy Majorana lepton\cite{Ng}, with resulting
rates of the order $10^{-16}$ sec$^{-1}$. \ A computation has halso been
done using a relativistic quark model for mesons\cite{Abad}, which made it
possible to consider small neutrino masses. \ In that case, the resulting
rates are several orders of magnitude smaller.

Here we wish to look at the possible contribution to $K^{\pm }\rightarrow
\pi ^{\mp }\mu ^{\pm }\mu ^{\pm }$ that doubly charged Higgs bosons might
make, and see how they compare to the prvious calculations just mentioned. \
These doubly-charged Higgs bosons are present in exotic Higgs
representations and can have lepton-number-violating couplings to any pair
of leptons. \ They appear in many models, and to provide some focus we shall
concentrate on the phenomenology of a model with one $|Y|=0$ and one $|Y|=2$
triplet\cite{Gunion2}. \ The features of this model that are relevant for
our analysis are the following:

The lepton-number-violating coupling to left-handed leptons is specified by
the following Lagrangian: 
\begin{equation}
L=ih_{ij}\psi _{iL}^{T}C\tau _{2}\triangle \psi _{jL}+h.c.
\end{equation}

\noindent where $\psi _{iL}$ is the two-component leptonic doublet and the $%
\Delta $ is the 2$\times 2$ representation of the $Y=2$ complex triplet. \
These couplings lead directly to Majorana masses for the neutrinos, so this
model does not rely on right-handed lepton partners in order to generate
neutrino masses. \ The Lagrangian in eq. (1) implies a Majorana mass term
for the neutrino given by 
\begin{equation}
m_{ij}=\frac{h_{ij}s_{H}\upsilon }{\sqrt{2}}
\end{equation}

\noindent where the subscript $ij$ refers to the lepton families, $\upsilon $
is related to the neutral Higgs vacuum expectation values, and $s_{H}$ is
the sine of a doublet-triplet mixing angle. \ 

For the case of the electron neutrino, the experimental observation of
nuclear double beta decay puts severe limits on the value of $m_{\nu _{e}}$
and as a result on $h_{ee}$. One obtains\cite{Gunion2} 
\begin{equation}
s_{H}h_{ee}\lesssim 6\times 10^{-12}.  \label{limit1}
\end{equation}

\noindent Clearly for the electron-electron case we would not expect the
model to predict any significant phenomenological consequences. \ However,
in general for some models\cite{Gunion1} there are certain niceties that
follow from the assumption that the expectation value of the neutral Higgs
vanishes. \ In that case the limit on $m_{\nu _{e}}$ is not relevant, and
from Bhabba scattering one may derive the limit\cite{Gunion2}\cite{Gunion1}%
\cite{Swartz} 
\begin{equation}
h_{ee}^{2}<10^{-5}GeV^{-2}\times m_{\Delta ^{++}}^{2}.  \label{limit2}
\end{equation}
\noindent Constraints on $h_{\mu \mu }$ can be obtained from the
experimental limit on muonium to antimuonium conversion, with the result 
\begin{equation}
h_{ee}h_{\mu \mu }<6\times 10^{-5}GeV^{-2}\times m_{\Delta ^{++}}^{2}.
\label{limit3}
\end{equation}

The physical states can be classified according to their transformation
properties, and one finds a fiveplet $H_{5}$ (which includes the
doubly-charged components), a threeplet $H_{3}$, and two singlets. \ There
are a variety of couplings of the Higgs bosons to fermions and to gauge
vector bosons. \ The ones in which we are potentially interested are 
\[
H_{5}^{++}W^{-}W^{-}\sim i\sqrt{2}gm_{W}s_{H}g_{\mu \nu } 
\]
\begin{equation}
H_{5}^{++}H_{3}^{-}W^{-}\sim -i\sqrt{2}c_{H}e(p_{++}-p_{-})_{\mu }
\label{couplings}
\end{equation}
\[
H_{3}^{+}q^{\prime }\bar{q}\sim \frac{gs_{H}}{2\sqrt{2}m_{W}c_{H}}%
[m_{q^{\prime }}(1+\gamma _{5})-m_{q}(1-\gamma _{5})]. 
\]

Finally, we remark that in the limit $s_{H}\rightarrow 0$ the standard model
is regained, with the singlet $H_{1}^{0}$ playing the role of the standard
Higgs boson with the standard couplings to fermions and vector mesons.

\section{k$^{+}\rightarrow \pi ^{-}\mu ^{+}\mu ^{+}$}

The quark-level diagrams that contribute to the decay amplitude are shown in
Fig. 1. \ For this particular model only the phenomenology of left-handed
fields was developed\cite{Gunion2}, but in any case limits on right-handed
currents would make them irrelevant here. \ We adopt the usual form for the
hadronic charged weak current 
\begin{eqnarray}
J^{\alpha } &=&\cos \theta _{c}[\bar{u}_{u}\gamma ^{\alpha }(1-\gamma
_{5})u_{d}+\bar{u}_{c}\gamma ^{\alpha }(1-\gamma ^{5})u_{s}]  \nonumber \\
&&+\sin \theta _{c}[\bar{u}_{u}\gamma ^{\alpha }(1-\gamma ^{5})u_{s}+\bar{u}%
_{c}\gamma ^{\alpha }(1-\gamma _{5})u_{d}]
\end{eqnarray}

\noindent The amplitude for Fig. 1(a) is then given by 
\begin{eqnarray}
T_{1a} &=&\frac{ig}{2\sqrt{2}}<\pi |J^{\mu }|0>\frac{i}{m_{W}^{2}}i\sqrt{2}%
gm_{W}s_{H}\frac{i}{m_{\Delta ^{++}}^{2}}  \nonumber \\
\times h_{\mu \mu } &<&leptons>\frac{i}{m_{W}^{2}}\frac{ig}{2\sqrt{2}}%
<0|J^{\mu }|k>
\end{eqnarray}
where $<leptons>=\bar{u}(l_{1})(1+\gamma
^{5})u^{c}(l_{2})-(l_{1}\leftrightarrow l_{2})$. \ As usual, the current
matrix elements can be expressed as $<0|J^{\mu }|k>=if_{k}p_{k}$ and $<\pi
\left| J^{\mu }\right| 0>=if_{\pi }p_{\pi }$, where $f_{\pi }$ and $f_{k}$
are the meson decay constants. \ The amplitude then reduces to 
\begin{equation}
T_{1a}=0.6G_{F}^{3/2}m_{k}f_{\pi }f_{k}E_{\pi }<leptons>\frac{s_{H}h_{\mu
\mu }}{m_{\Delta ^{++}}^{2}}.  \label{T1a}
\end{equation}

The amplitude for Fig. 1(b) cannot be written down in a similar simple form.
\ In order to estimate its contribution we begin by writing down the
amplitude at the quark level: 
\begin{eqnarray}
A &=&\frac{ig}{2\sqrt{2}}\cos \theta _{c}\bar{u}_{d}\gamma ^{\alpha
}(1-\gamma ^{5})u_{u}\frac{i}{m_{W}^{2}}i\sqrt{2}gm_{W}s_{H}\frac{i}{%
m_{\Delta ^{++}}^{2}}  \label{A} \\
\times h_{\mu \mu } &<&leptons>\frac{i}{m_{W}^{2}}\frac{ig}{2\sqrt{2}}\sin
\theta _{c}\bar{v}_{s}\gamma ^{\alpha }(1-\gamma ^{5})v_{u}.  \nonumber
\end{eqnarray}

\noindent We project out the ''pseudoscalar'' content of $A$ with the
operation\cite{Richardson} 
\begin{equation}
T_{1b}=C_{\pi }C_{k}\sum \bar{u}_{u}\gamma ^{5}v_{s}A\bar{v}_{u}\gamma
^{5}u_{d}
\end{equation}

\noindent where the sum is over the quark spins and $C_{\pi }$ and $C_{k}$
are normalization constants. \ The amplitude is then evaluated using a
simple static constituent quark model. \ The decay constants are determined
by applying the procedure to the decays $K\rightarrow l\nu $ and $\pi
\rightarrow l\nu $. \ The result yields $C_{k}=if_{k}m_{k}/8m_{s}m_{u}$ and $%
C_{\pi }=if_{\pi }m_{\pi }/8m_{u}m_{d}$. \ The method can be tested by
applying it to the amplitude of Fig. 1(a) and calculating the decay rate for
that amplitude alone. \ In that case the result is the same as that found
from the earlier calculation for the amplitude given in Eq. (\ref{T1a}).

Applying this procedure to Eq. (\ref{A}) yields 
\begin{equation}
T_{1b}=-1.2\times 10^{3}G_{F}^{3/2}C_{\pi }C_{k}m_{\pi }^{2}m_{k}E_{\pi
}<leptons>\frac{s_{H}h_{\mu \mu }}{m_{\Delta ^{++}}^{2}}.
\end{equation}

The remaining diagrams in Fig. 1 can be evaluated in similar fashion. \ For
example, $T_{1e}$ is given by 
\begin{eqnarray}
T_{1e} &=&i\frac{g^{3}h_{\mu \mu }}{2\sqrt{2}}\sin \theta _{W}\sin \theta
_{c}\cos \theta _{c}\frac{s_{H}}{m_{W}^{3}m_{\Delta ^{+}}^{2}m_{\Delta
^{++}}^{2}}<leptons>  \nonumber \\
&&\times (m_{s}-m_{u})\bar{v}_{s}^{\alpha }(1+\gamma ^{5})u_{u}\bar{u}%
_{d}\gamma ^{\alpha }(1-\gamma ^{5})v_{u}(p_{++}+p_{-})^{\alpha }.
\end{eqnarray}

\noindent where $p_{++}$ is the momentum of the incoming $\Delta ^{++}$ and $%
p_{-}$ the momentum of the outgoing $\Delta _{-}$.

It is clear that with reasonable values for the Higgs masses the
contributions from these graphs will be considerably smaller than that from
the first two. \ Therefore we calculate the rate using only the diagrams in
Fig 1(a,b). \ After squaring the amplitude and doing the spin sum we
integrate over phase space as follows: 
\begin{equation}
R=\frac{1.1\times 10^{-18}}{4(2\pi )^{3}}\frac{1}{m_{k}^{3}}\left( \frac{%
s_{H}h_{\mu \mu }}{m_{\Delta ^{++}}^{2}}\right) ^{2}\int ds\lambda
^{1/2}(s,m_{k}^{2},m_{\pi }^{2})\lambda ^{1/2}(s,m_{\mu \mu }^{2},m_{\mu \mu
}^{2})E_{\pi }^{2}p_{1}\cdot p_{2}.
\end{equation}

\noindent Here $s=(p_{1}+p_{2})^{2}$, $p_{1}$ and $p_{2}$ are the muon
momenta, and as usual $\lambda (x,y,z)=x^{2}+y^{2}+z^{2}-2xy-2xz-2yz.$ \
Evaluation of the integral yields the following result for the rate: 
\begin{equation}
R=6\times 10^{-2}\left( \frac{s_{H}h_{\mu \mu }}{m_{\Delta ^{++}}^{2}}%
\right) ^{2}\text{GeV}^{4}\text{ \ }\frac{1}{\sec }.  \label{rate}
\end{equation}

\section{RESULTS AND DISCUSSION}

It is possible to develop models that incorporate Higgs triplets where the $%
\Delta ^{++}W^{-}W^{-}$coupling is allowed only with an enabling non-zero
vacuum expectation value for the neutral member\cite{Gunion1}. \ In that
case one may obtain a limit for $h_{ee}$ through the Majorana mass term in
neutrinoless nuclear double beta decay, similar to that shown in Eq. (\ref
{limit1}). \ The assumption that the vev is zero avoids problems with
maintaining $\rho =1$ , but in that case graphs like Figs. 1(a,b) would not
be allowed. \ However, couplings of the triplet to vector bosons of the type
shown in Eq. (\ref{couplings}) are a natural consequence of the model being
discussed here. \ 

At the same time, the coupling of Higgs to fermions is determined by the
overlap of the mass-eigenstate Higgs fields with the doublet, and the $H_{5}$
does not have such overlap. \ \ The $H_{3}$ does and that makes graphs in
Figs. 1(c-h) possible, but the contributions to the rate are smaller by at
least a factor $\sim (m_{k}/m_{\Delta ^{+}})^{4}$. \ 

It is clear then that the rate will be dominated by Figs. 1(a,b), with the
result shown in Eq. (\ref{rate}) and with a branching ratio given by 
\begin{equation}
BR=0.9\times 10^{-9}\left( \frac{s_{H}h_{\mu \mu }}{m_{\Delta ^{++}}^{2}}%
\right) ^{2}\text{GeV}^{4}.
\end{equation}

\noindent The existing experimental limit\cite{Data} for this branching
ratio is $1.5\times 10^{-4}$, corresponding to 
\begin{equation}
s_{H}h_{\mu \mu }\leq 1.5\times 10^{2}(m_{\Delta ^{++}}/\text{GeV})^{2}.
\end{equation}

\noindent This result is not invalidated by the previous limits in Eqs.
(\thinspace \ref{limit1},\ref{limit2},\ref{limit3}), but it is not exactly a
giant step forward either. \ Taking the values $m_{\Delta ^{++}}\sim 100$ GeV%
\cite{Gunion1} and $s_{H}h_{\mu \mu }\sim 5\times 10^{-2}m_{\Delta ^{++}}$
(near to and not excluded by the limits of Eqs. (\ref{limit2},\ref{limit3}))
yields a branching ratio of $2\times 10^{-16}$, which is discouragingly
small but of the same order of magnitude as the rate for kaon double beta
decay induced by light or heavy Majorana neutrinos and reported in earlier
publications\cite{Ng}\cite{Abad}.

In conclusion, we have calculated the rate for $K^{\pm }\rightarrow \pi
^{\mp }\mu ^{\pm }\mu ^{\pm }$ that can be expected for a decay mediated by
doubly-charged Higgs. \ The branching ratio for reasonable values of the
model's parameters is extremely small compared to the experimental limit,
but of the same order of magnitude as that induced by virtual Majorana
neutrinos. Since the question of lepton-number violation is still
outstanding, these processes warrant further theoretical and experimental
study.

This work was supported in part by the Natural Sciences and Engineering
Research Council of Canada.

\begin{center}
\bigskip Figure Captions
\end{center}

\noindent FIG. 1. \ Diagrams contributing to $K^{+}\rightarrow \pi ^{-}\mu
^{+}\mu ^{+}$.

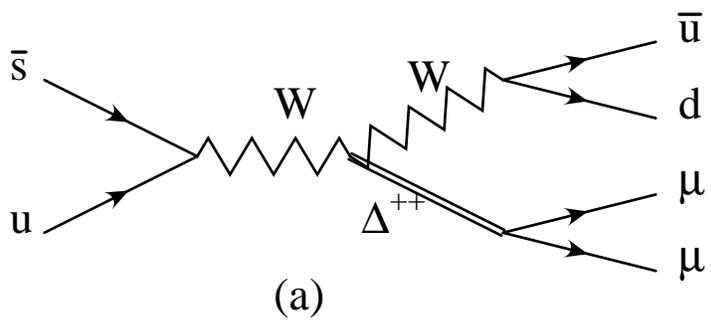
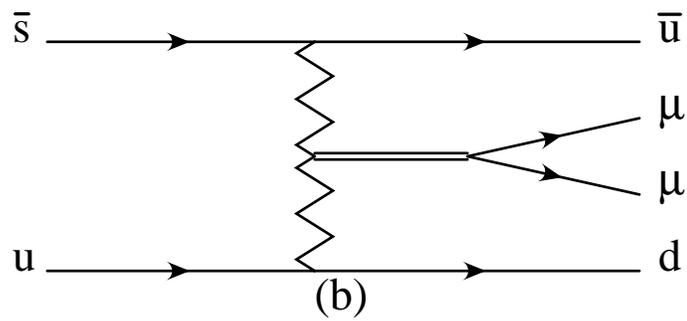
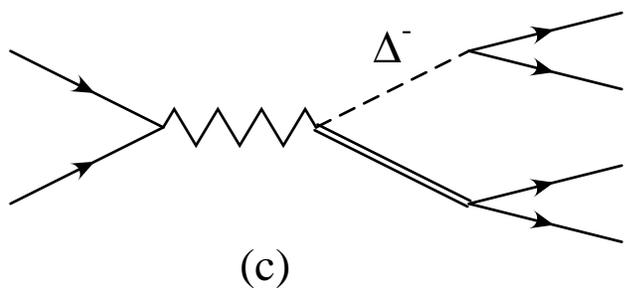
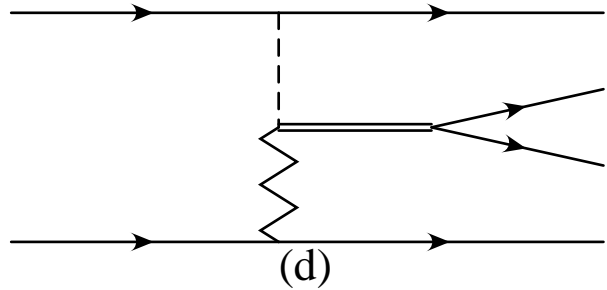
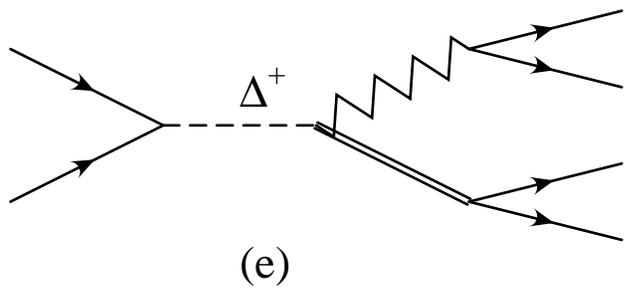
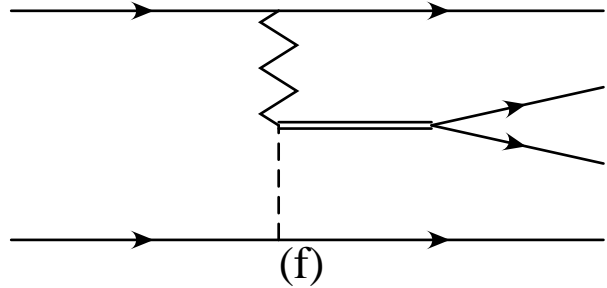
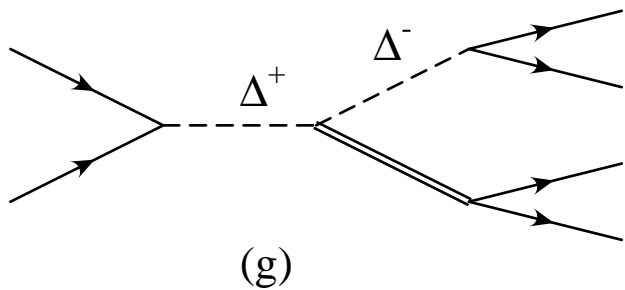
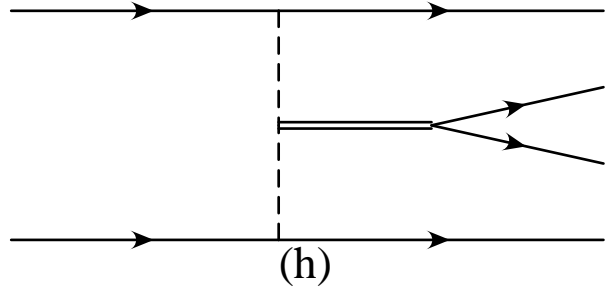
\end{document}